\documentclass[aip,reprint]{revtex4-1}

\usepackage{graphicx}
\usepackage{subfigure}
\usepackage{natbib}
\usepackage{amsmath,amssymb}

\draft 

\begin{document}


\title{Hierarchical sinuous-antenna phased array for millimeter wavelengths} 



\author{Ari Cukierman}
\email[]{ajcukierman@berkeley.edu}
\affiliation{Department of Physics, University of California, Berkeley, CA 94720, USA}

\author{Adrian T. Lee}
\affiliation{Department of Physics, University of California, Berkeley, CA 94720, USA}
\affiliation{Physics Division, Lawrence Berkeley National Laboratory, Berkeley, CA 94720, USA}
\affiliation{Radio Astronomy Laboratory, University of California, Berkeley, CA 94720, USA}

\author{Christopher Raum}

\affiliation{Radio Astronomy Laboratory, University of California, Berkeley, CA 94720, USA}

\author{Aritoki Suzuki}

\affiliation{Physics Division, Lawrence Berkeley National Laboratory, Berkeley, CA 94720, USA}

\author{Benjamin Westbrook}

\affiliation{Radio Astronomy Laboratory, University of California, Berkeley, CA 94720, USA}


\date{\today}

\begin{abstract}
We present the design, fabrication and measured performance of a hierarchical sinuous-antenna phased array coupled to superconducting transition-edge-sensor (TES) bolometers for millimeter wavelengths. The architecture allows for dual-polarization wideband sensitivity with a beam width that is approximately frequency-independent. We report on measurements of a prototype device, which uses three levels of triangular phased arrays to synthesize beams that are approximately constant in width across three frequency bands covering a 3:1 bandwidth. The array element is a lens-coupled sinuous antenna. The device consists of an array of hemispherical lenses coupled to a lithographed wafer, which integrates TESs, planar sinuous antennas and microwave circuitry including band-defining filters. The approximately frequency-independent beam widths improve coupling to telescope optics and keep the the sensitivity of an experiment close to optimal across a broad frequency range. The design can be straightforwardly modified for use with non-TES lithographed cryogenic detectors such as kinetic inductance detectors (KIDs). Additionally, we report on the design and measurements of a broadband $180^\circ$ hybrid that can simplify the design of future multichroic focal planes including but not limited to hierarchical phased arrays.
\end{abstract}


\maketitle 


Lithographed superconducting detectors have found many applications in millimeter and submillimeter astronomy including measurements of the cosmic microwave background (CMB), galactic dust emission and star and planet formation.\cite{Zmuidzinas2004}
A major goal of modern cosmology is to make precise measurements of the polarized fluctuations in the CMB with science goals such as a measurement of the sum of neutrino masses and a detection of primordial $B$-modes, which would be considered direct evidence for an inflationary epoch in the early universe.\cite{Seljak1996_GravityWaves,CMBS4Science2017} The signals of interest are at the level of $10$-$100~\mathrm{nK}$ on a $2.7$-$\mathrm{K}$ background, which motivates the development of increasingly sensitive instruments.\cite{Benson2014,Suzuki2016,Matsumura2016,Grayson2016,Thornton2016,Essinger-Hileman2014,CMBS4Technology2017} In this letter, we propose an efficient method for improving the sensitivity of such experiments.

The noise-equivalent temperature ($\mathrm{NET}$), often given in units of $\mu\mathrm{K} \sqrt{\mathrm{s}}$, is a measure of the instantaneous sensitivity of an experiment to a sky signal such as the CMB. The noise in an experiment\rq{}s maps of the sky can be reduced by integrating for longer lengths of time. The mapping speed $\mathrm{MS} = 1/\mathrm{NET}^2$, often given in units of $\mathrm{\mu K}^{-2}~\mathrm{s}^{-1}$, is a measure of the rate at which the map noise is reduced. An experiment with twice the mapping speed can achieve the same map noise in half the time.

\begin{figure}
	\includegraphics[width = 0.35\textwidth]{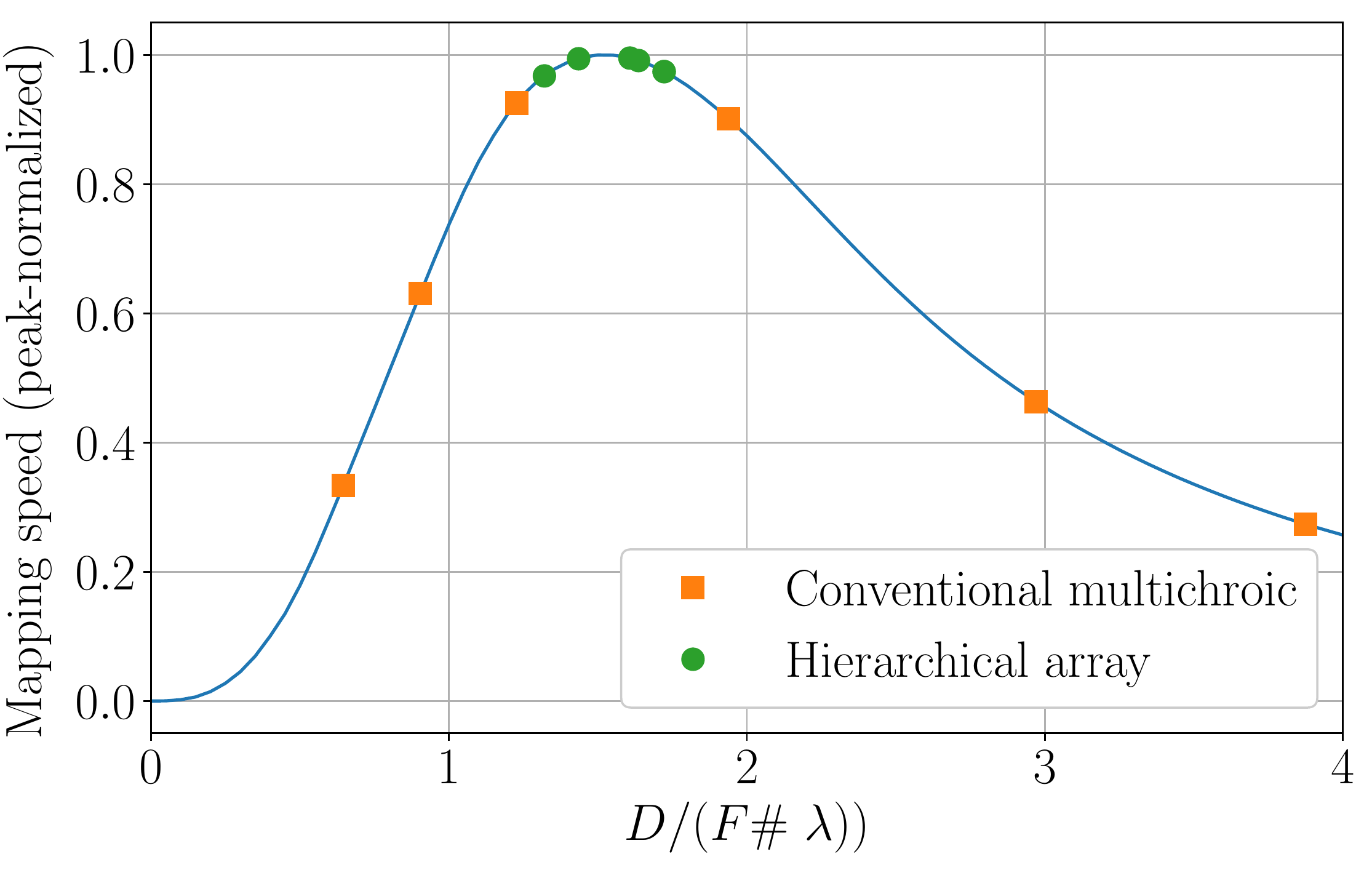}
	\caption{Mapping speed vs. pixel diameter (\emph{solid}) in units of f-number ($F\#$) of the detector-aperture system times wavelength ($\lambda$) for an example multichroic focal plane for a space-based telescope\cite{Matsumura2016}: $2.7$-$\mathrm{K}$ sky with a $4$-$\mathrm{K}$ aperture stop and bands centered on $50$, $70$, $95$, $150$, $230$ and $300~\mathrm{GHz}$. A more precise treatment would yield slightly different curves for different frequencies, but the optimal diameter typically varies by less than $10\%$ in units of $F\#~\lambda$. \emph{Square} markers indicate where the frequency bands of a conventional multichroic pixel, i.e., with one fixed diffracting aperture, would fall on this curve when the pixel diameter is optimized for $120~\mathrm{GHz}$; \emph{circular} markers indicate the same for a hierarchical focal plane based on triangular arrays for which the array-element diameter is optimized for $270~\mathrm{GHz}$. The hierarchical array allows near-optimal mapping speed in all frequency bands. \label{fig:mappingSpeed}}
\end{figure}

The dependence of the mapping speed on pixel size is well known in the CMB community.\cite{Griffin2002} Here we consider the common case of single-moded detectors and a cryogenic aperture stop with the angular size of the telescope's field of view held constant. A smaller pixel size allows for more pixels, which tends to increase the mapping speed. However, a smaller pixel has a wider beam and, therefore, receives a smaller fraction of power from the sky and a larger fraction from the aperture stop in the telescope optics, which contributes only noise. There are, therefore, two competing concerns in choosing a pixel size. A plot of mapping speed vs. pixel size for an example experiment is shown in Fig.~\ref{fig:mappingSpeed}, which shows a clear peak indicating an optimal pixel diameter. 
The pixel can be treated roughly as a diffraction aperture, so the diameter-to-wavelength ratio sets the beam width. As a result, the optimal pixel diameter is approximately proportional to wavelength. 

A common problem with a conventional multichroic pixel, which is sensitive to multiple frequency bands but has a single fixed pixel diameter, is that the beam width and, therefore, the optimal pixel diameter are frequency-dependent. It is usually impossible, then, to choose a pixel diameter that is optimal in every frequency band. A \emph{multiscale} focal plane solves this problem by creating pixels whose effective diameters scale with wavelength.

\begin{figure}
	\includegraphics[height = 0.15\textwidth]{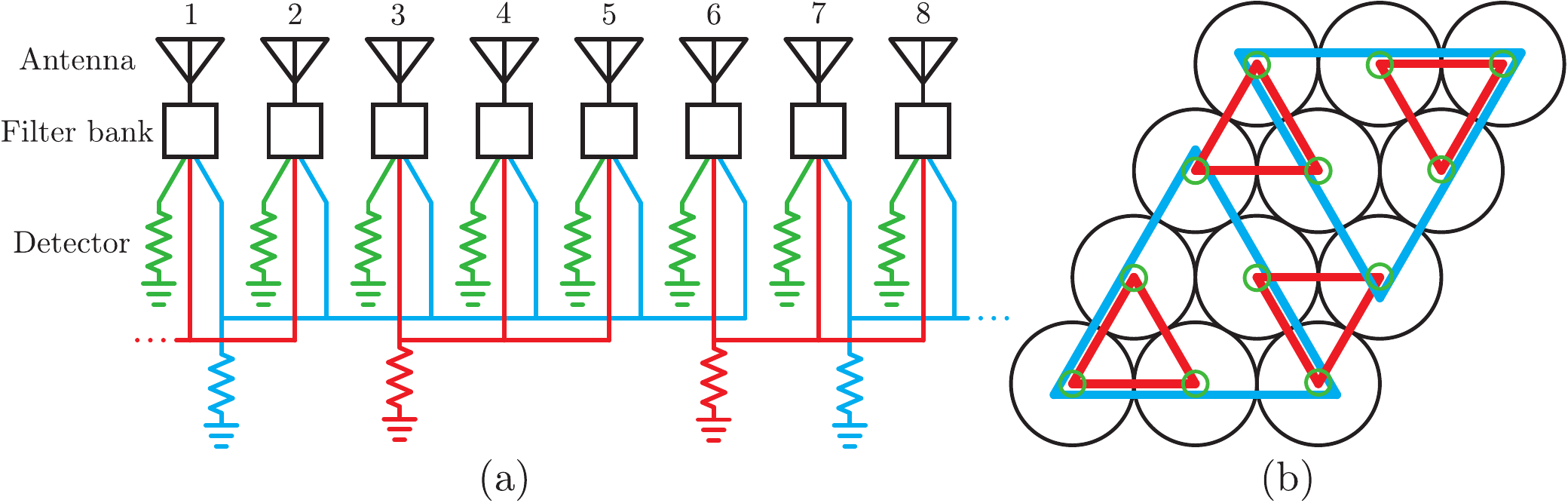}
	\caption{(\emph{a}) Example topology of a three-level hierarchical phased array.  The high frequencies (\emph{green}) dissipate power to bolometers in a conventional way. The middle frequencies (\emph{red}) sum three antennas coherently before dissipating power. The low frequencies (\emph{blue}) sum six antennas. (\emph{b}) Effective pixels for a three-level hierarchy based on triangular arrays. \emph{Black circles} represent individual antennas. The \emph{small green circles} represent the effective pixels for the high frequencies, \emph{small red triangles}  the middle frequencies  and \emph{large blue triangles}  the low frequencies. The triangular hierarchy naturally creates a second hierarchy of rhombi, which can be used to tile the wafer.  \label{fig:topology}}
\end{figure}

A multiscale focal plane can be realized through the use of a \emph{hierarchical phased array}, which consists of superimposed antenna arrays operating in different frequency bands and increasing in size approximately inversely with frequency. An example topology is shown in Fig.~\ref{fig:topology} for a three-level hierarchy based on triangular arrays.
The antennas couple free-space radiation to transmission lines, at which point the power is split into frequency bands via the filter banks. Transmission-line signals from neighboring antennas are then summed coherently before dissipating the power on a bolometer. The summing network includes a different number of antenna elements for each frequency band. The estimated improvement to mapping speed is shown in Fig.~\ref{fig:mappingSpeed}.

An additional benefit of a hierarchical focal plane is that the detector count per unit area may be reduced relative to a conventional design, even though the sensitivity may be improved. This is possible when the conventional design is optimized for high frequencies. In this case, the hierarchical design simply combines antennas at lower frequencies, which reduces the number of detectors per antenna. Whereas the conventional design would have a detector count that increases approximately linearly with the number of frequency bands, the hierarchical design has a detector count that increases approximately logarithmically. Under these conditions, a three-band experiment could reduce its detector count by half.

This paradigm is particularly attractive for satellite- and balloon-based experiments, for which increased weight and power are expensive. Additionally, the mapping speed tends to have a narrower optimum due to the larger optical loading from the aperture stop relative to the sky in the absence of atmospheric loading. For ground-based experiments, the reduction in system complexity, coming, e.g., from decreased power consumption and wire count for readout electronics, is also attractive.

A hierarchical focal plane can also mitigate systematic errors related to beam shape and polarization purity. The far-field beam is the product of the array factor and the beam of the individual antenna element. Beam non-idealities can be compensated for by choosing an appropriate array configuration. Some antennas, including the sinuous antenna discussed below, have a polarization angle that varies or ``wobbles'' with frequency.\cite{Edwards2012} By constructing an array consisting partially of mirror-imaged versions of the antenna element, the polarization wobble can be reduced or even cancelled completely.

\begin{figure}
	\includegraphics[width = 0.3\textwidth]{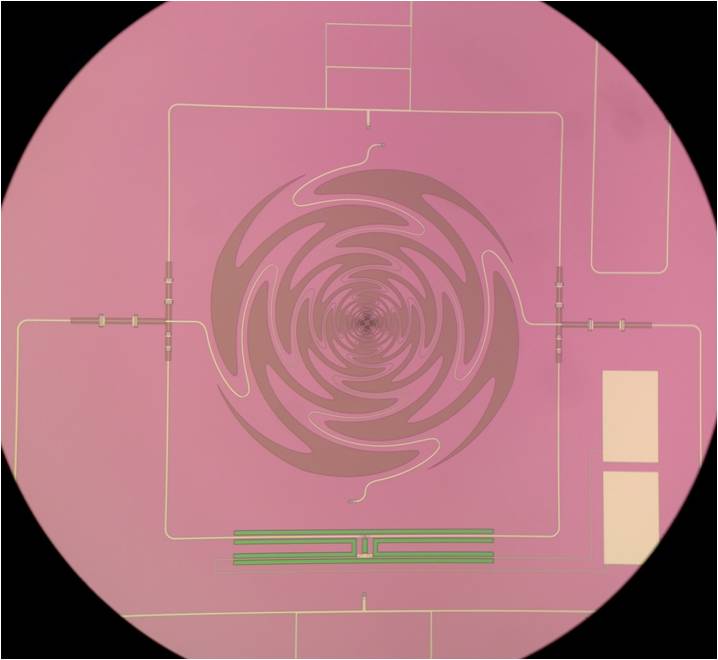}
	\caption{Unit cell of the hierarchical array: a sinuous antenna coupled to out-of-phase microstrip lines, triplexing bandpass filters (\emph{middle, left and right}) and double-ring hybrids (\emph{top and bottom}). The $220$-$\mathrm{GHz}$ signal is sent directly to a bolometer (\emph{bottom center}). For simplicity of fabrication, only the polarization associated with the horizontal sinuous microstrip lines is active; the vertical lines are terminated. \label{fig:pixel}}
\end{figure}

We have developed a multiscale architecture based on hierarchical phased arrays that allows for dual-polarized wideband sensitivity at millimeter wavelengths.
As the unit cell of these arrays, we choose the lenslet-coupled sinuous antenna on account of its broad bandwidth, frequency-independent beam waist and dual-polarized capability.\cite{Edwards2012,OBrient2013,Westbrook2016} Since the beam waist is determined mainly by the lens diameter, the on-chip planar antenna need only occupy a fraction of the total pixel area, which makes space for microwave circuitry and detectors.  The on-chip components of the unit cell is shown in Fig.~\ref{fig:pixel}. 
We find that the beam forming of the antenna arrays compensates for beam non-idealities in the unit cell, so the sinuous antenna can be substantially undersized relative to the case of a conventional multichroic pixel; here we use a $1.7$-$\mathrm{mm}$ diameter (cf. 3-mm for a conventional pixel operating at the same frequencies\cite{Benson2014}). 

To simplify and accelerate the fabrication process for this prototype, we chose to avoid microstrip crossovers, since this would have added extra layers. This prototype is, consequently, a single-polarization device, although the architecture can be extended, by adding well-proven microstrip crossovers, to two polarizations. Separately, we have been developing microstrip cross\emph{under}s, which perform the same function as crossovers but without requiring additional layers. We plan to make use of these crossunders in future hierarchical devices.

\begin{figure}
	\includegraphics[height = 0.2\textwidth]{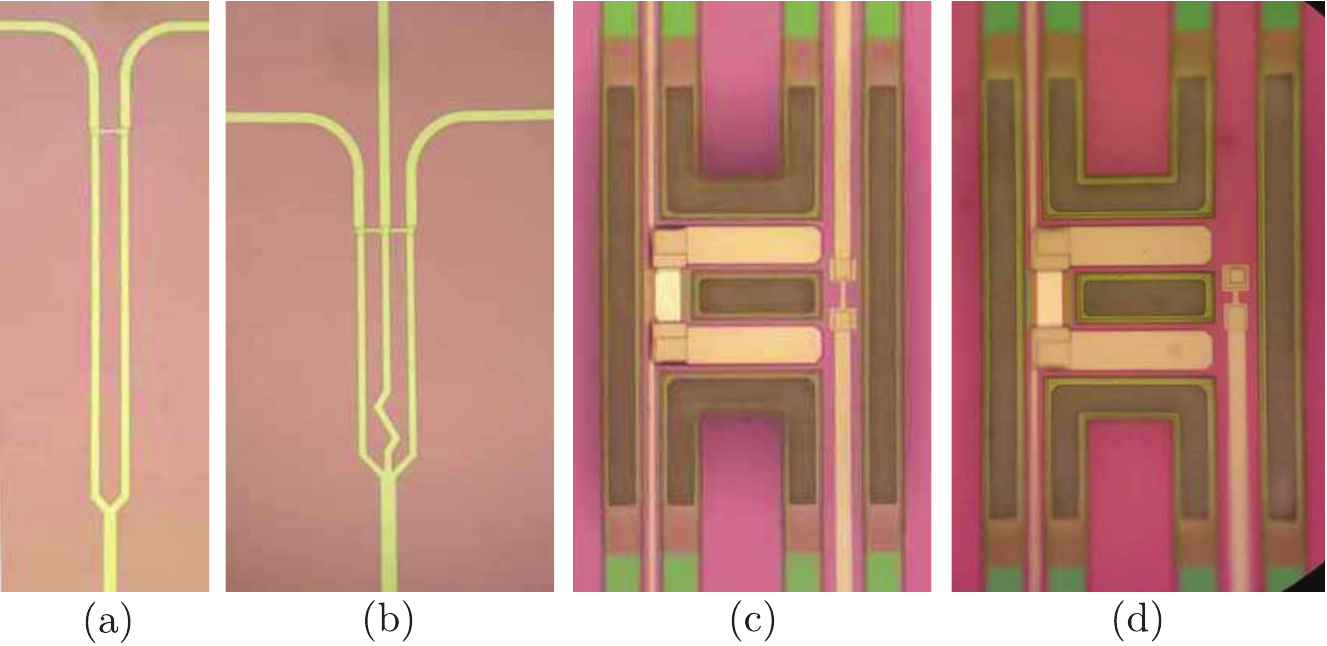}
	\caption{(\emph{a})~A Wilkinson divider consisting of superconducting $\mathrm{Nb}$ microstrip lines and a $\mathrm{Ti}$ resistor (connecting the output lines near the top). (\emph{b})~A planarized Wilkinson ``trivider''. (\emph{c})~A bolometer island featuring an $\mathrm{AlMn}$ TES (\emph{left middle}) and microstrip lines terminating differentially on a $\mathrm{Ti}$ resistor (\emph{right middle}). (\emph{d})~A bolometer island featuring a single microstrip line terminating in a $\mathrm{Ti}$ resistor that is connected through a via in the insulating $\mathrm{Si} \mathrm{O}$ to the $\mathrm{Nb}$ ground plane. \label{fig:uwaveComponents}}
\end{figure}

The sinuous antenna is coupled to two out-of-phase microstrip lines that route the signal to triplexing lumped-element bandpass filters, which split the signal into bands centered on $90$, $150$ and $220~\mathrm{GHz}$ with bandwidths of approximately~$25\%$. The $220$-$\mathrm{GHz}$ signal is brought directly to a bolometer, where the microstrip lines are terminated differentially on a lumped resistor (Fig.~\ref{fig:uwaveComponents}(c)). The out-of-phase lines for $90$ and $150~\mathrm{GHz}$ are combined in double-ring hybrids to form single microstrip lines, which enter the summing networks.

\begin{figure}
	\includegraphics[width = 0.3\textwidth]{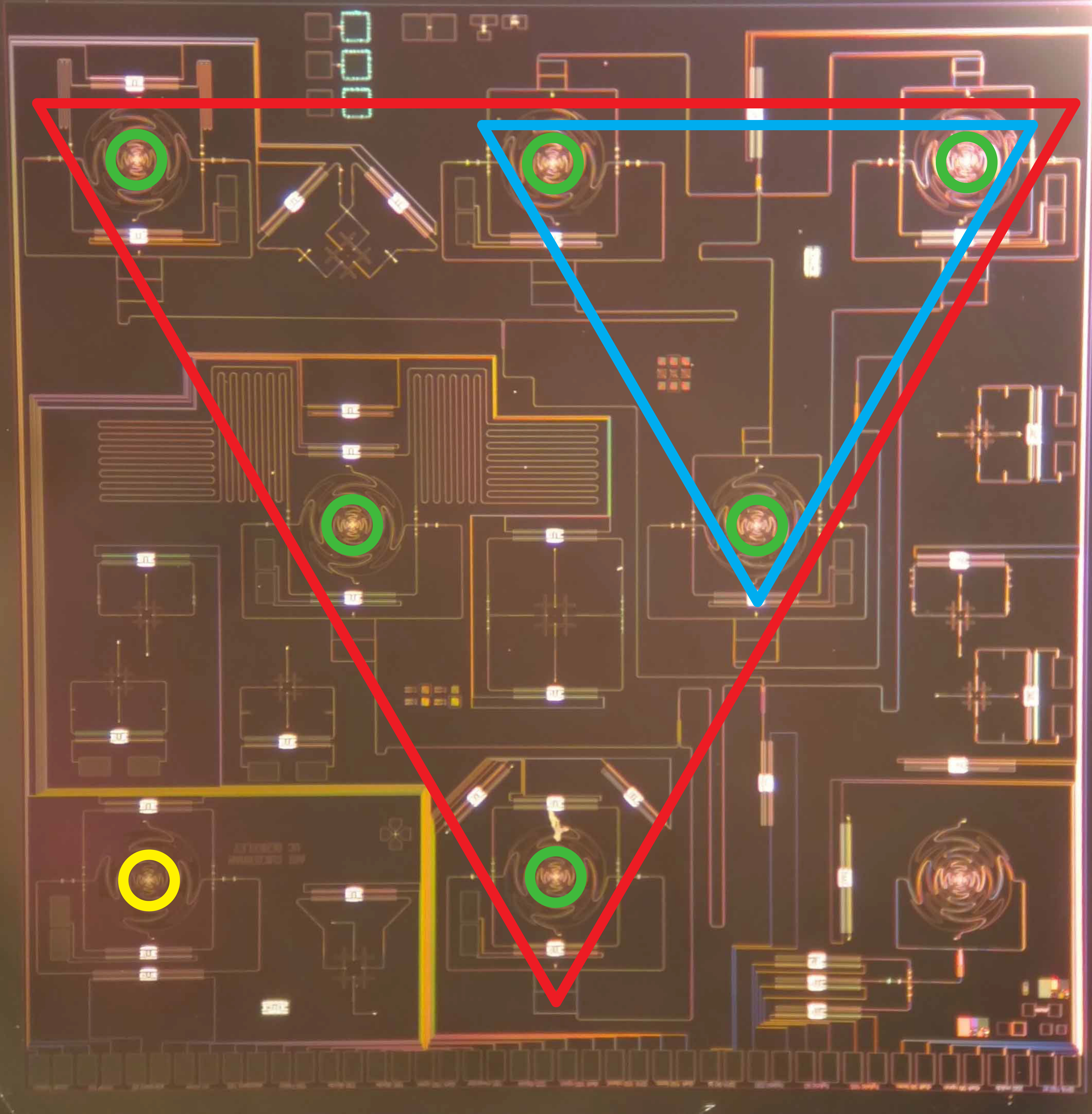}
	\caption{Fabricated prototype chip with overlaid hierarchical structure: $220$-$\mathrm{GHz}$ pixels (\emph{green}), $150$-$\mathrm{GHz}$ array (\emph{blue}) and $90$-$\mathrm{GHz}$ array (\emph{red}). A conventional trichroic pixel (\emph{yellow}) is included as a control. The other features on the chip are for test pixels that were included in the same mask design but are electrically independent of the hierarchical array. \label{fig:chip}}
\end{figure}

An overview of the summing networks is shown in Fig.~\ref{fig:chip}. 
The signals from neighboring antennas are combined with the Wilkinson power splitters\cite{Wilkinson1960} shown in Fig.~\ref{fig:uwaveComponents}. 
An advantage of these splitters is that the output ports are well-isolated from each other, which suppresses unwanted reflections between antennas. To increase their bandwidths, these Wilkinson splitters include additional $\lambda/4$ segments ahead of the splitting node (only partially shown in Fig.~\ref{fig:uwaveComponents}). To planarize the Wilkinson ``trivider'', we removed the resistor that would have connected the outermost output lines and then retuned the circuit parameters in simulation. The 3-antenna $150$-$\mathrm{GHz}$ array is formed using a single trivider. The 6-antenna $90$-$\mathrm{GHz}$ array is formed using three dividers followed by a single trivider. The microstrip lengths between the antennas and the splitters are identical within each frequency band to avoid phase delays that steer the beam off-zenith. 

The summed signal is terminated on a lumped resistor on a bolometer island (Fig.~\ref{fig:uwaveComponents}(d)), where the signal is detected by an $\mathrm{Al} \mathrm{Mn}$ TES with a critical temperature of approximately $450~\mathrm{mK}$. The fabrication process and bolometer design are similar to those developed for the POLARBEAR-2 (PB-2) detector wafers.\cite{Westbrook2016,Suzuki2016} 

For the microstrip ground plane and strip layer, we use superconducting $\mathrm{Nb}$. As a microstrip dielectric, we used $\mathrm{Si} \mathrm{O}$, but we intend to use $\mathrm{Si} \mathrm{N}$ for future devices, since it shows $\sim 10$ times lower loss and has been implemented in the wafers recently fabricated for PB-2.

The wafer was mated with a lenslet array\cite{QuealyThesis} and installed in a test cryostat, which cools the detector wafer to $270~\mathrm{mK}$. The bolometers are read out with DC superconducting quantum interference devices (SQUIDs).

\begin{figure}
	\includegraphics[width = 0.35\textwidth]{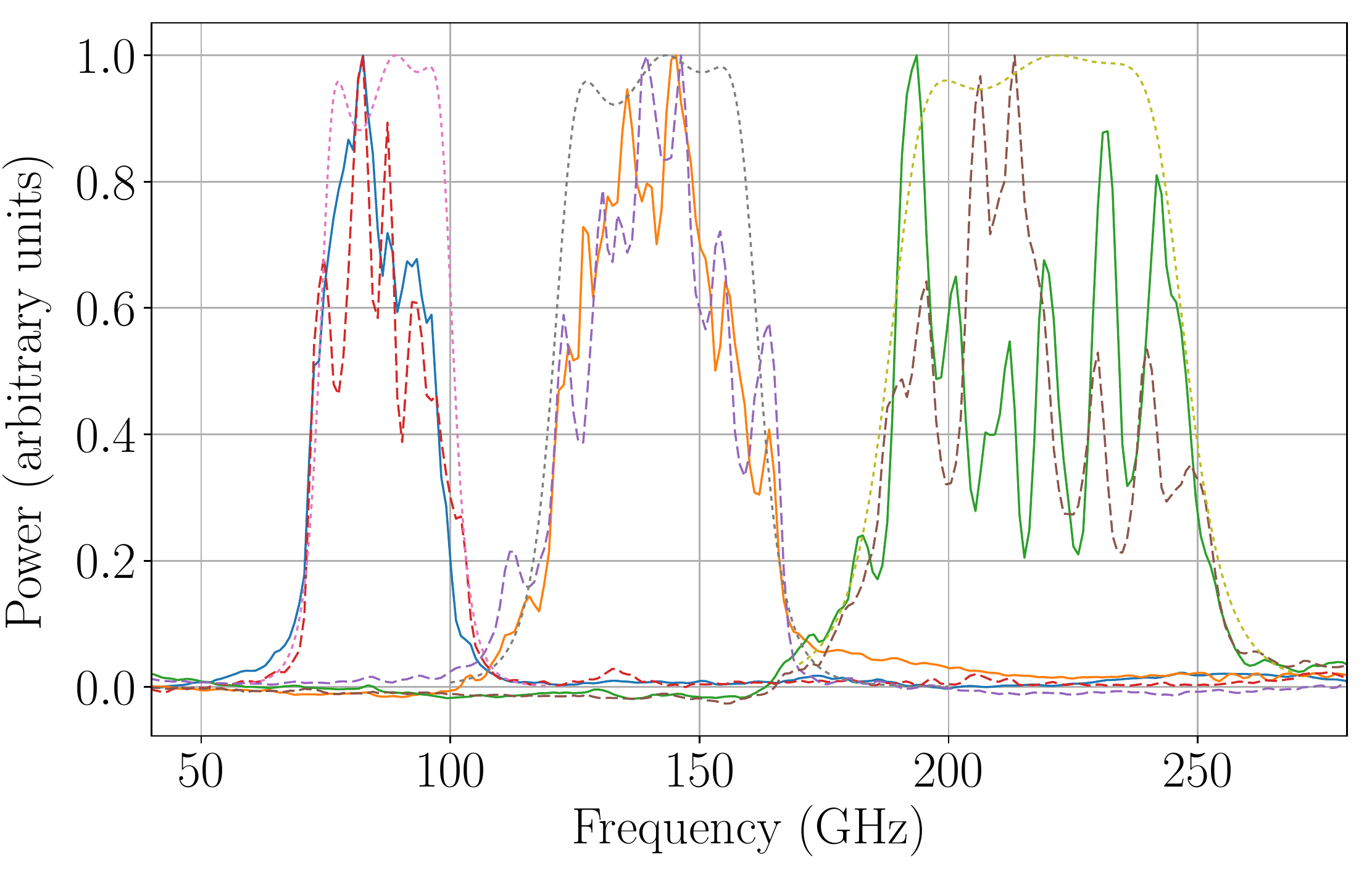}
	\caption{The frequency response shows bands centered near $90$, $150$ and $220~\mathrm{GHz}$ for both the conventional trichroic pixel (\emph{dashed}) and the hierarchical array (\emph{solid}), which show similar band edges to the simulated bandpass filters (\emph{dotted}). \label{fig:FTS}}
\end{figure}

The frequency response was measured with a Fourier-transform spectrometer. The results (Fig.~\ref{fig:FTS}) are similar for the conventional and hierarchical pixels, showing that the hierarchical design is not introducing a spurious frequency response.
Relative to the design, the measured bands shifted down in frequency by $5$-$10\%$. This is common for a first prototype and is likely due to an incorrect dielectric constant or dielectric thickness in simulation. The shift can be tuned away by varying the dielectric thickness on subsequent iterations. The simulated bands shown in Fig.~\ref{fig:FTS} use a $15\%$ thinner microstrip dielectric than originally intended. A similar shift can be achieved by increasing the dielectric constant by~$15\%$. 

\begin{figure}
	\includegraphics[height = 0.36\textwidth]{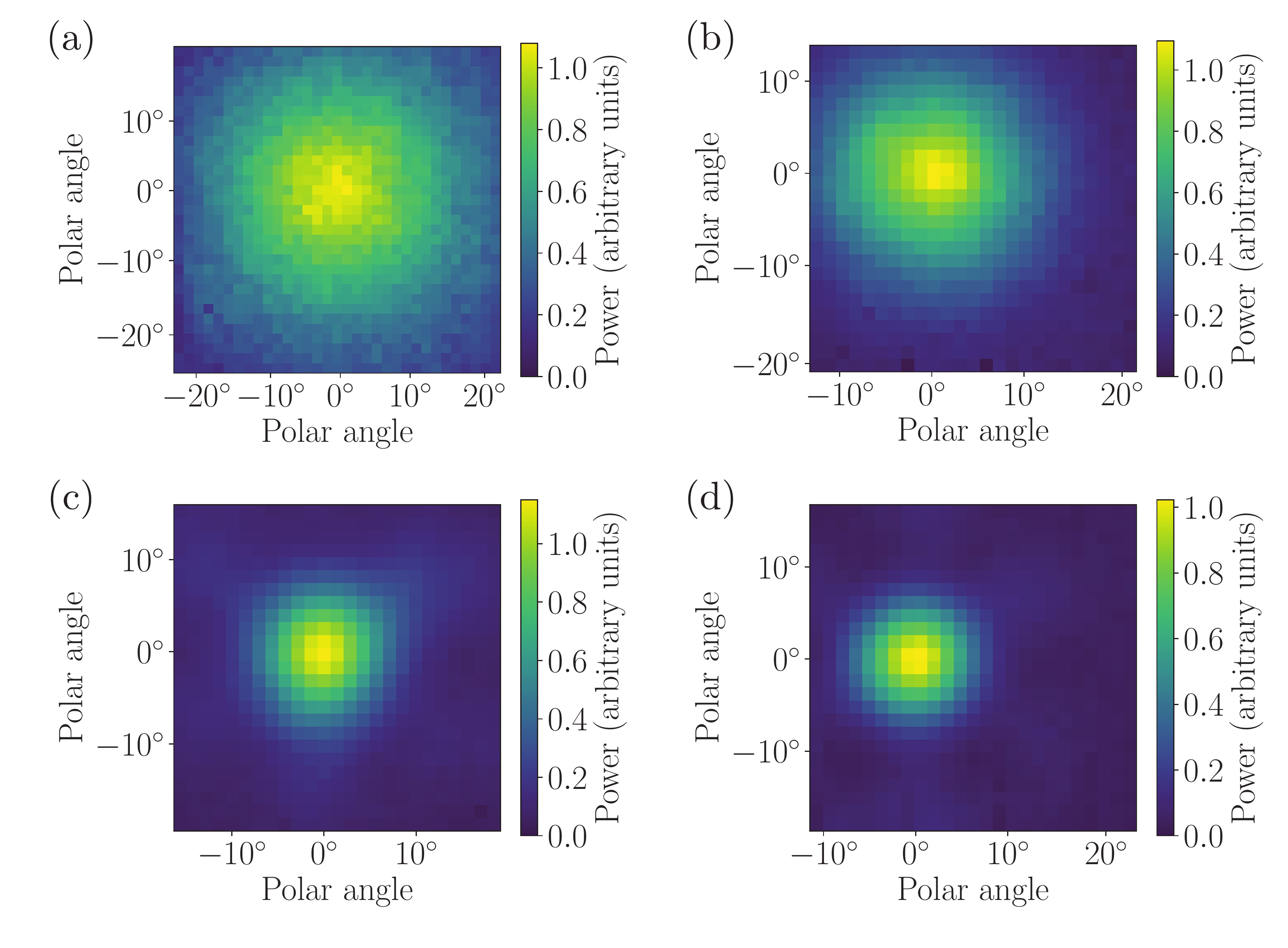}
	\caption{Beam maps for the $90$-~(\emph{a}) and $150$-$\mathrm{GHz}$~(\emph{b}) conventional channels and the $90$-~(\emph{c}) and $150$-$\mathrm{GHz}$~(\emph{d}) hierarchical-array channels. 
The conventional trichroic pixel shows frequency-dependent beam widths. The hierarchical array shows roughly constant beam widths. \label{fig:beamMaps}}
\end{figure}

\begin{figure}
	\includegraphics[width = 0.35\textwidth]{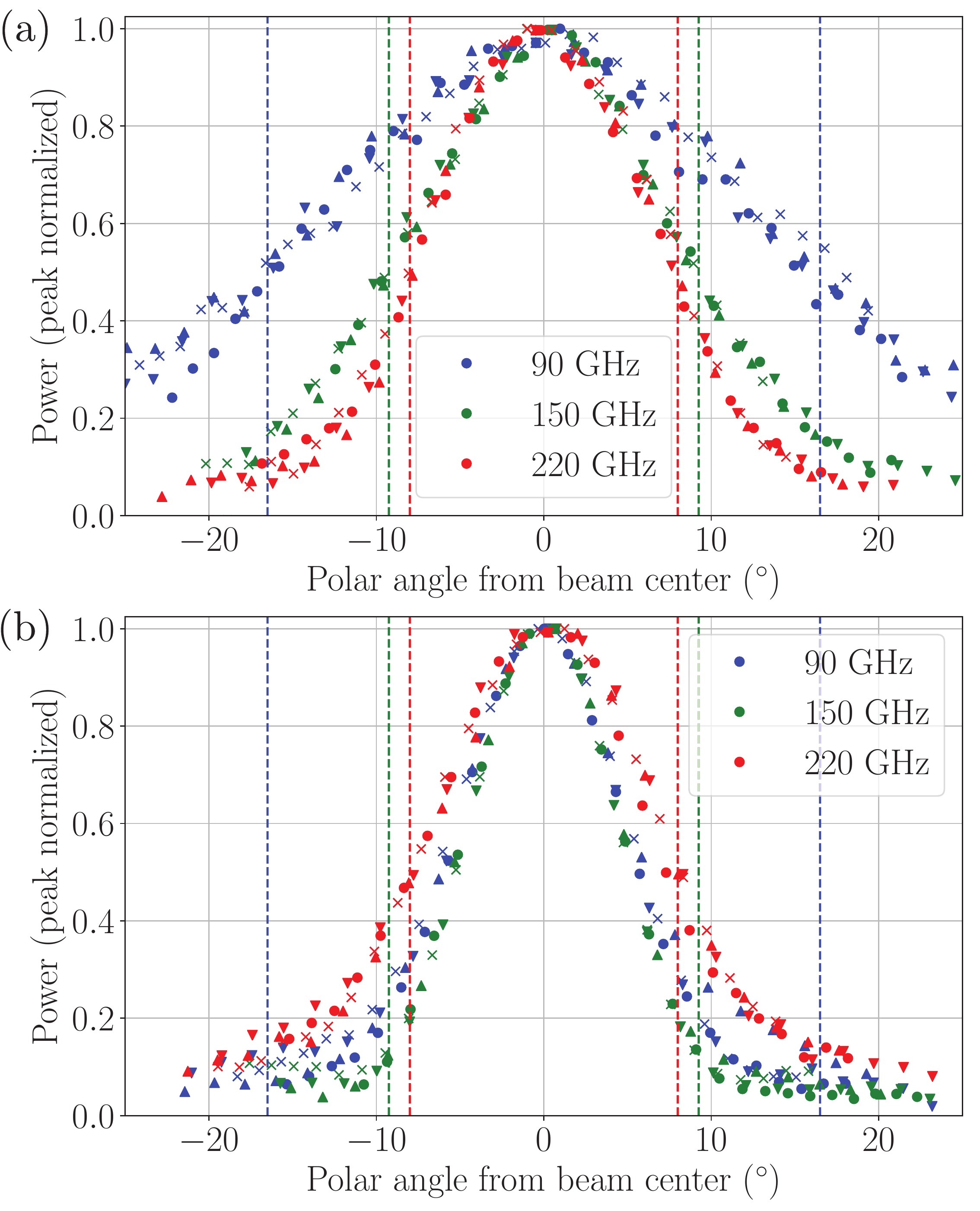}
	\caption{Beam profiles for the conventional trichroic pixel~(\emph{a}) and the hierarchical array~(\emph{b}). The different markers correspond to different azimuthal angles, i.e., $0^\circ$, $45^\circ$, $90^\circ$ and $135^\circ$. To guide the eye, the vertical \emph{dashed} lines indicate the half maxima for the conventional pixel. \label{fig:beamCuts}}
\end{figure}

The beams were measured using a chopped liquid-nitrogen source on an XY stage. The results are shown in Fig.~\ref{fig:beamMaps} for $90$ and $150~\mathrm{GHz}$.
The array has created round beams of approximately the same size at both frequencies with widths that are within $15\%$ of the expectations from electromagnetic simulations for the $90$- and $150$-$\mathrm{GHz}$ bands. The $220$-$\mathrm{GHz}$ beams were wider than the expectations by $20$-$25\%$, which may be due to imperfections in the lenslet and anti-reflection coating that affect smaller wavelengths more strongly. The hierarchical array reduces the beam width by a factor of~$2.6$ at~$90$ and $1.8$ at~$150~\mathrm{GHz}$ compared with expectations of~$2.9$ and~$2.1$, respectively. The faint hexagonal pattern is expected from the array factor of triangular antenna arrays. The beam profiles are shown in Fig.~\ref{fig:beamCuts}, where we see that the hierarchical design has created approximately constant beam widths across the $90$, $150$ and $220$-$\mathrm{GHz}$ bands.

We were unable to measure the optical efficiency, since the optical loading from the room-temperature equipment exceeded the saturation power of the TESs. The measurements were performed, therefore, above the superconducting transition. Non-linearities in the detector response were removed in post-processing. We place conservative lower bounds on the optical efficiency relative to top-hat bands of fractional width $25\%$ centered on $90$, $150$ and $220~\mathrm{GHz}$. We estimate $> 50\%$ optical efficiency at $90~\mathrm{GHz}$, $> 30\%$ at $150~\mathrm{GHz}$ and $>25\%$ at $220~\mathrm{GHz}$.

\begin{figure}
	\includegraphics[width = 0.3\textwidth]{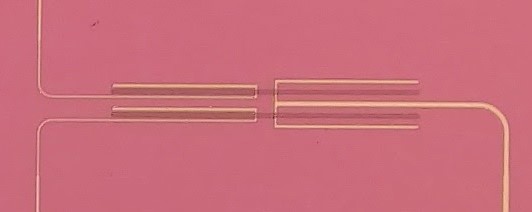}
	\caption{A broadband $180^\circ$ hybrid relying on microstrip-slotline coupling. The two lines at left couple asymmetrically to the two slotlines; the line at right couples symmetrically. A signal injected at right is split evenly into the two left lines but with a $180^\circ$ phase shift between them. The two lines at left feed the sinuous antenna differentially. The line at right enters bandpass filters that split the signal into frequency bands. \label{fig:broadbandHybridPixel}}
\end{figure}

\begin{figure}
	\includegraphics[height = 0.14\textwidth]{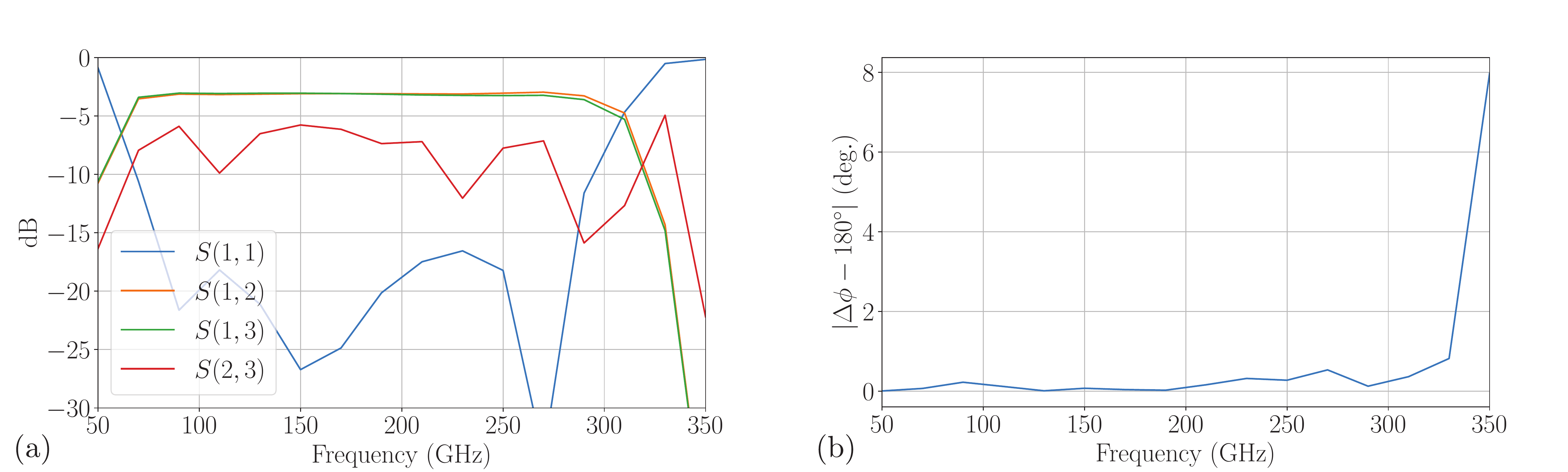}
	\caption{Simulations of the broadband $180^\circ$ hybrid. (\emph{a})~$S$-parameters showing roughly even power splitting ($S(1,2) \approx S(1,3) \approx - 3 ~\mathrm{dB}$) from $70$ to $280~\mathrm{GHz}$. (\emph{b})~Phase difference of the output ports relative to~$180^\circ$. \label{fig:broadbandHybridSim} }
\end{figure}

\begin{figure}
	\includegraphics[height = 0.14\textwidth]{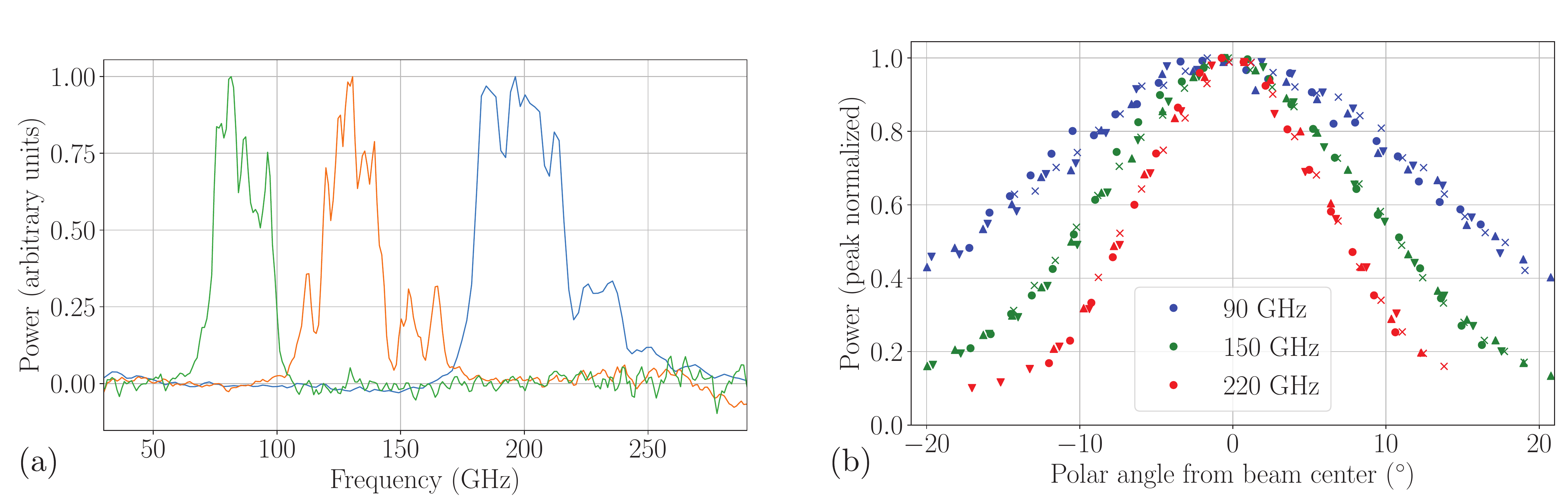}
	\caption{Measurements of a trichroic pixel using a broadband $180^\circ$ hybrid. (\emph{a})~Frequency response. (\emph{b})~Beam profiles. \label{fig:broadbandHybridMeasurements}}
\end{figure}

Looking forward to future hierarchical designs, we have been developing the lenslet-coupled sinuous antenna to cover a broad frequency range from $40$ to $350~\mathrm{GHz}$\cite{Westbrook2016} as well as testing microwave components such as a broadband hybrid. As a potential replacement for the relatively narrowband double-ring hybrids (Fig.~\ref{fig:pixel}), we designed a $180^\circ$ hybrid that operates across a 4:1 bandwidth. This component saves space, simplifies the design and reduces the number of microstrip cross-overs. Our design (Fig.~\ref{fig:broadbandHybridPixel}) relies on a microstrip-slotline-microstrip coupling. 
To test this design, we built a prototype pixel that is similar to the conventional trichroic pixel described above but uses a broadband hybrid to combine the antenna feedlines before entering the bandpass filters. Unlike the conventional trichroic pixel, the broadband-hybrid pixel only has one set of bandpass filters instead of two. Simulation results are shown in Fig.~\ref{fig:broadbandHybridSim}.
Measurements of the bands  (Fig.~\ref{fig:broadbandHybridMeasurements}) show some degradation relative to those of Fig.~\ref{fig:FTS} but are correctly placed.
The beams are all co-centered to better than~$1^\circ$. The lack of frequency-dependent beam steering indicates that the hybrid splits power evenly with a $180^\circ$ phase difference over the entire frequency range. 

\begin{figure}
	\includegraphics[width = 0.35\textwidth]{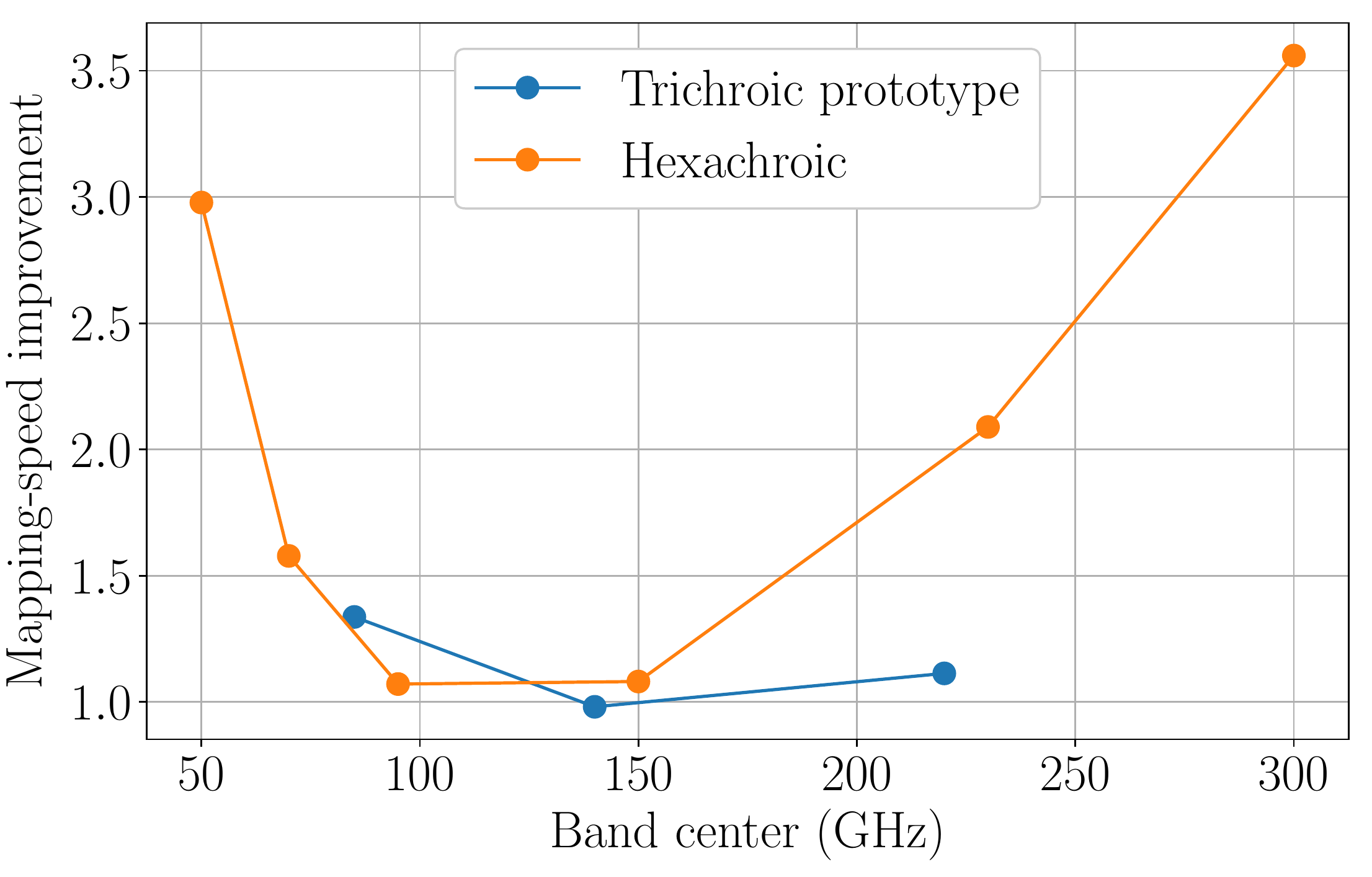}
	\caption{Mapping-speed improvement factor from hierarchical phased arrays for the prototype described in this letter, where the effective pixel size has been chosen to be optimal at~$120~\mathrm{GHz}$, and for the ultrawideband design described in Fig.~\ref{fig:mappingSpeed}. The improvement becomes more significant for larger total bandwidths. The prototype produces the relatively modest improvements of $10\%$ at $220$ and $30\%$ at $90~\mathrm{GHz}$. \label{fig:mappingSpeedImprovement} }
\end{figure}

We have demonstrated the essential features of a hierarchical sinuous phased array, which has the potential to substantially improve the mapping speed of multichroic focal planes (Fig.~\ref{fig:mappingSpeedImprovement}).
The gains are most significant for relatively large total bandwidth.
The architecture does not depend on detector type and can be straightforwardly adapted for non-TES lithographed cryogenic detectors such as kinetic inductance detectors (KIDs). We envision this technology as an integral part of future wideband millimeter-wave telescopes. \\

This work was supported by NASA grant NNX17AH13G. The devices were fabricated in the Marvell Nanolab at UC Berkeley.


%
%

%



\bibliography{Cukierman}

\end{document}